\documentclass[3p,times]{elsarticle}

\usepackage{graphicx} 
\usepackage{siunitx}
\usepackage{amsmath}
\usepackage{bm}
\usepackage{xurl}
\usepackage{hyperref}
\hypersetup{breaklinks=true}
\usepackage[version=4]{mhchem}
\usepackage{orcidlink}

\newcommand\bfe{\mathbf{e}}
\newcommand\bfn{\mathbf{n}}
\newcommand\bfx{\mathbf{x}}
\newcommand\bfy{\mathbf{y}}
\newcommand\bfF{\mathbf{F}}
\newcommand\bfM{\mathbf{M}}
\newcommand\bfS{\mathbf{S}}
\newcommand\emi{\varepsilon}
\begin{document}

\sloppy
\raggedright

\begin{frontmatter}
\title{Determination of emissivity profiles using a Bayesian data-driven approach}

\author[1]{Luca Sgheri \orcidlink{0000-0001-5734-0856}}
\ead{luca.sgheri@cnr.it}
\author[2]{Cristina Sgattoni \orcidlink{0000-0001-5734-0856}}
\ead{cristina.sgattoni@cnr.it}
\author[3,1]{Chiara Zugarini \orcidlink{0009-0005-8053-0646}}
\ead{chiara.zugarini@fi.iac.cnr.it}

\affiliation[1] {organization={CNR-IAC},
                 addressline={Via Madonna del Piano, 10},
                 city={Sesto Fiorentino},
                 postcode={I-50019},
                 state={FI},
                 country={Italy}
}
\affiliation[2] {organization={CNR-IBE},
                 addressline={Via Madonna del Piano, 10},
                 city={Sesto Fiorentino},
                 postcode={I-50019},
                 state={FI},
                 country={Italy}
}
\affiliation[3] {organization={Università di Firenze, Dipartimento di Ingegneria Informatica},
                 addressline={Via di S. Marta, 3},
                 city={Firenze},
                 postcode={I-50139},
                 state={FI},
                 country={Italy}
}

\begin{abstract}
In this paper, we explore the determination of a spectral emissivity profile that closely matches real data, intended for use as an initial guess and/or a-priori information in a retrieval code. Our approach employs a Bayesian method that integrates the CAMEL (Combined ASTER MODIS Emissivity over Land) emissivity database with a land cover map. The solution is derived as a convex combination of high-resolution Huang profiles using the Bayesian framework. We test our method on IASI (Infrared Atmospheric Sounding Interferometer) data and find that it outperforms the linear spline interpolation of the CAMEL data.
\end{abstract}

\end{frontmatter}
\section{Introduction and background}
Any body hit by electromagnetic radiation either absorbs or reflects the radiation. The absorbed radiation is then re-emitted as thermal radiation according to the body temperature and Planck's law. The fraction $\emi$ of radiation absorbed by the body is called emissivity, and can vary from $0$ (all radiation reflected) to $1$ (all radiation absorbed). The remaining fraction $r$ is called reflectivity and by the energy conservation law $\emi + r = 1$.

Earth is the body that the remote sensing community is interested in. Earth emissivity depends on the wavelength of the electromagnetic radiation, and it is often termed spectral emissivity. It also depends on the kind of surface, so there is a dependence from latitude and longitude. This also engenders a partial seasonal dependence. Barren land and cultivated fields have different emissivity profiles, and snow covering alters emissivity.

Radiative transfer \cite{chandrasekhar1960} is the equation that models the intensity of electromagnetic radiation as it travels through a not empty mean. Interferometers that are monitoring the atmosphere from space measure the spectrum, i.e. the intensity of radiation as a function of wavelength, as it is seen by the instrument.

We will focus on instruments measuring in the infra-red spectral band. More precisely, we will consider the Far-infrared Outgoing Radiation Understanding and Monitoring (FORUM) mission \cite{palchetti2018,palchetti2020,sgheri2022}, which will be the 9th Earth-Explorer mission of the European Space Agency and will sound the atmosphere in the Far Infra-Red (FIR) to Middle Infra-Red (MIR) interval $100$-$1600$~\si{cm^{-1}} from $2027$.

If the atmospheric state is known, the spectrum can be modeled using the radiative transfer equation along the line of sight of the instrument. To obtain an estimate of the atmospheric composition, the radiative transfer equation must be inverted. The inversion of the radiative transfer is also called retrieval. 

The radiative transfer for a layer of a homogeneous mean and a single wavelength can be derived from the Lambert-Beer and Planck laws (see for instance \cite{zdunkowski2007}) and can be written:
\begin{equation}\label{eq:rt1}
\left\{\begin{array}{l}
    \frac{d}{dz}I(z)=-\alpha I(z) + \alpha B(T) \\
    I(z_0)={I}_0,
    \end{array}\right.
\end{equation}
where $z$ is the spatial coordinate, $I(z)$ is the intensity of the electromagnetic radiation, $T$ is the temperature of the layer, $B$ is the Planck function and $\alpha$ is the attenuation coefficient, that depends on pressure, temperature and concentrations of the various atmospheric gases in the layer. Equation (\ref{eq:rt1}) has the analytical solution
\begin{equation}\label{eq:rt_sol}
I(z)={I}_0\exp(-\alpha(z-z_0))+B(T)(1-\exp(-\alpha(z-z_0))).
\end{equation}

For a nadir-looking spectrometer mounted on a satellite the radiative transfer equation extends from Earth level $z_0$ to the top of atmosphere $z_N$, an altitude of about $120$~\si{km}. Above this altitude the density is so low that we can consider the space to be void. If the atmosphere is discretized in homogeneous layers $[z_{i-1},z_i]$, with $T_i$ the Curtis-Godson \cite{godson1953} average temperature of the layer $i$ and $\alpha_i$ the attenuation coefficient of the layer, the optical depth of the layer is $\tau_i=\alpha_i(z_i-z_{i-1})$. Then, if $\emi$ and $T_E$ are the emissivity and temperature of Earth surface at the location of the scan, respectively, the solution of the radiative transfer can be written \cite{zdunkowski2007}:
\begin{equation}\label{eq:RTE_discr}
\begin{split}
I(z_{N})=&\left[ \epsilon \ B(T_{E}) + (1-\epsilon) \left( \sum_{i=1}^{N} B(T_{i}) \ (1 - e^{-\tau_{i}}) \ e^{-\sum_{j=1}^{i-1} \tau_{j}} \right) \right] e^{-\sum_{i=1}^{N} \tau_{i}}\\
&+\sum_{i=1}^{N} B(T_{i}) \ (1 - e^{-\tau_{i}}) \ e^{-\sum_{j=i+1}^{N} \tau_{j}}. 
\end{split}
\end{equation}
In the retrieval inverse problem under clear sky conditions the unknowns are the surface temperature and emissivity, as well as the vertical profiles of temperature and concentration of the main atmospheric gases. The temperature and concentrations in (\ref{eq:RTE_discr}) are embedded in the attenuation coefficients $\alpha_i$, but in this paper we only focus on emissivity. Note, however, that having unknowns in the combination of exponentials lead to a severely ill-conditioned problem, see for instance \cite{rodgers2000,transtrum2011}.

One of the most common methods for inverting the radiative transfer is the Optimal Estimation (OE) method by Rodgers \cite{rodgers2000}, that combines in a Bayesian approach two mathematical techniques.
\begin{enumerate}
    \item The use of a weighted $L^2$ norm for calculating the error between the instrumental and simulated spectra;
    \item The use of a Tikhonov penalization term, using the weighted $L^2$ norm of the difference between the retrieved profiles and some a-priori.
\end{enumerate}

Let $\bfx$ be the set of parameters of the retrieval problem, $\bfF$ the forward model, i.e. the radiative transfer, and $\bfy$ the spectrum measured by the instrument. If we suppose that the $\bfy$ measurements are random variables with a Gaussian experimental error, we can define the Variance-Covariance Matrix (VCM) of the measurements as
\begin{equation}\label{VCM}
\bfS_y=\mathbb{E}((\bfy-\mathbb{E}(\bfy))(\bfy-\mathbb{E}(\bfy))^t),
\end{equation}
where $\mathbb{E}$ is the expected value. If we have a model $\bfx_a$ for the parameters $\bfx$ and we know the variability of the a-priori, we can also define the VCM $\bfS_a$ of the a-priori in the same way. The a-priori $\bfx_a$ may come for instance from the climatology or from correlated measurements. The OE method consists in minimizing the following functional:
\begin{equation}\label{eq:oe}
\chi^{2}(\bfx) = (\bfy-\bfF(\bfx))^{t}\bfS_{y}^{-1}(\bfy-\bfF(\bfx)) \, + \, (\bfx_{a}-\bfx)^{t}\bfS_{a}^{-1}(\bfx_{a}-\bfx).
\end{equation}
The calculation of the attenuation coefficients is a heavy computational task, because it involves obtaining information from a spectroscopic database such as the HIgh resolution TRANsmission (HITRAN) 2020 database, see \cite{gordon2022}. On the other hand, the calculation of the Jacobian of $\bfF$ with respect to $\bfx$ is easily obtained because of the exponential dependence. Thus, the minimization is normally carried out using an analytical method such as the Gauss-Newton (GN) method with the Levemberg-Marquardt technique to counter the non-linearity of the problem.

There are several issues in the retrieval of spectral emissivity. First, there are bands where the absorption lines of atmospheric gases mask the signal from Earth. For instance, in the $300$-$600$~\si{cm^{-1}} interval the absorption lines of water vapour make the atmosphere opaque unless we are at polar latitudes, where the atmosphere is very dry~\cite{harries2008}. On the other hand, in the MIR region we have two atmospheric windows, roughly corresponding to $800$-$950$~\si{cm^{-1}} and $1020$-$1200$~\si{cm^{-1}} intervals. In these frequency ranges the atmosphere in clear-sky conditions is highly transparent because there are very few absorption lines, so that surface parameters (temperature and emissivity) can be retrieved. For a recent study of the emissivity sensitivity in the FORUM spectral range see \cite{sgattoni2024}.

Secondly, in the radiative transfer (\ref{eq:RTE_discr}) the dependence of the spectrum $I(z_N)$ on $\emi$ is linear. In the atmospheric windows, where there is the maximal sensitivity to emissivity, the Planck function can also be approximated with a straight line with negative slope, since the maximum of the Planck function for attainable surface temperatures is reached around $600$~\si{cm^{-1}}. Thus, also the dependence of the spectrum on the surface temperature is almost linear, and this leads to negative covariances between  the two quantities \cite{li2013,benyami2022}. The anti-correlation can reach values of $-0.8$ in the atmospheric windows \cite{sgheri2022}. Note that this is only a correlation linked to the radiative transfer equation, and does not imply any physical correlation between surface temperature and emissivity.

There are different way of tackling this problem, to avoid biases in the retrieved surface temperature and spectral emissivity. We list here some techniques.
\begin{enumerate}
    \item From Equation (\ref{eq:RTE_discr}) we note that, while $B(T_E)$ does not depend on the wavelength, we could define a different $\emi$ for each measurement. Naturally, due to the experimental errors, the emissivity profile cannot be reconstructed pointwise, and a piecewise linear spline is instead commonly used. A coarser grid reduces the random noise, since a single value averages more spectral measurements, but fine features of real emissivity cannot be reproduced. On the other hand a finer grid has a larger random noise, but in principle can reproduce the small-scale features that are present in some emissivity profiles.
    \item Changing the $\bfS_a$ also influences the retrieval. Assigning a larger noise than expected permits larger variations of emissivity in the iterations of the GN sequence. Also, reducing or eliminating the correlations in the $\bfS_a$ permits a sharper definition of the emissivity profile in the transition zones of the sensitivity regions. However, a weaker regularization may trigger unwanted oscillations in the emissivity profile. An a-posteriori regularization such as the Iterative Variable Strength (IVS) \cite{ridolfi2011,sgheri2020} may be needed to reduce this phenomenon.
    \item While the limited variation of the emissivity permits to use a constant a-priori with typical values of $0.9$ or $0.99$, a more accurate prediction of the emissivity profile can be determined using the geolocation and time of the scan.
\end{enumerate}
The first two techniques are considered in \cite{sgheri2022} for FORUM retrievals, though a more complete study would be desirable, especially when the final specifications of the instrument are fully defined. In this paper we study the third issue, i.e. finding an a-priori statistically closer to the truth using emissivity models and climatological data.

\section{Emissivity models and climatologic datasets}

\subsection{Huang set of profiles}
There are different models for emissivity profiles as a function of the type of soil. The Huang profiles \cite{huang2016} are very detailed, cover the whole FORUM range, and have a resolution of $5$~\si{cm^{-1}}. In this paper we add a profile for tropical to middle latitude forests to the set of eleven Huang profiles used in the FORUM E2E simulator \cite{sgheri2022}. The soil types of the Huang profiles selected are reported in Table~\ref{ta:emissivity} of Section~\ref{sec:metodology}.

\subsection{Emissivity databases}
The Moderate Resolution Imaging Spectroradiometer (MODIS) database from the University of Wisconsin \cite{borbas2007,seemann2008} is available online from the site: \url{https://lpdaac.usgs.gov}. The MOD11 products consist in monthly means of emissivity at the following 10 wavelengths: $3.6$, $4.3$, $5.0$, $5.8$, $7.6$, $8.3$, $9.3$, $10.8$, $12.1$, and $14.3$~\si{\mu m} with $0.05^{\circ}$ spatial resolution. 
The Jet Propulsion Laboratory produces a Global Emissivity Dataset (GED) using the Advanced Spaceborne Thermal Emission and Reflection (ASTER) Radiometer \cite{aster,hulley2015}, available from \url{https://emissivity.jpl.nasa.gov/aster-ged}. The ASTER GEDv3 contains five emissivity values in the TIR band at multiple resolutions: 100~\si{m}, 1~\si{km} and 5~\si{km}.
These two databases were combined into The Combined ASTER MODIS Emissivity over Land (CAMEL) database \cite{borbas2018,feltz2018}.

All these databases rely on the concept of \emph{hinge points}, the points that should permit to obtain a good approximation of the emissivity profile using linear interpolation in between. Alternatively, a high resolution emissivity profile can be determined using principal components \cite{masiello2014}. We will use both the MODIS database and the CAMEL databases. In the MODIS database, $6$ out of the $10$ emissivity values are included in the FORUM spectral range, corresponding to: $669.30$, $884.96$, $943.40$, $1098.90$, $1162.79$ and $1204.82$~\unit{cm^{-1}} wavenumbers. In the CAMEL database, $9$ out of the $13$ emissivity values are included in the FORUM spectral range, corresponding to the MODIS values plus the $826.45$ $925.93$ and $1315.79$~\unit{cm^{-1}} wavenumbers.

\subsection{Atmospheric and ancillary data}
The ECMWF (European Center for Medium range Weather Forecast) database ERA5 \cite{hersbach20}, available at \url{https://www.ecmwf.int/en/forecasts/dataset/ecmwf-reanalysis-v5}, contains a source of atmospheric states, obtained with the assimilation of measurements from many different instruments. The ERA5 version contains a hourly database of geolocated vertical profiles of pressure, temperature, water vapour and ozone content on a $0.25^\circ$ grid in latitude and longitude. On the same grid surface temperature and pressure are available. Additional gas concentrations for \ce{CO2}, \ce{CO}, \ce{CH4}, \ce{NO2}, \ce{HNO3}, \ce{SO2} are available from the same source. Missing atmospheric gas profiles can be taken from the Initial Guess 2 (IG2) databases \cite{remedios2007}.

Ancillary data that will be used in our setting are the monthly snow cover fraction and soil humidity percentage \cite{peters-lidard2008}. While NRT maps are available, for instance from \url{https://www.nohrsc.noaa.gov/nh_snowcover/}, in this study we only use climatology. In the ERA5 database there is a soil type parameter which is linked to the soil texture. However, a very detailed map of soil types with a resolution of $0.05^\circ$~in latitude and longitude is available from the land cover MODIS data \url{https://doi.org/10.5067/MODIS/MCD12C1.006} \cite{Friedl2015}.

\section{Methodology}\label{sec:metodology}
Our key point is to make use of the high-resolution Huang emissivity profiles, that have proven to be accurate, while linking the selection to the available emissivity measurements and ancillary data. We propose two approaches, the first one is based on empirical constrains on ancillary data but no information about the soil type is used as an a-priori. In the second approach, the information about the soil type is mapped to the relevant point, using a Bayesian approach.

\subsection{Selection of profiles via ancillary data}
\begin{figure}[ht!]
\centering
\includegraphics[width=1.
\textwidth]{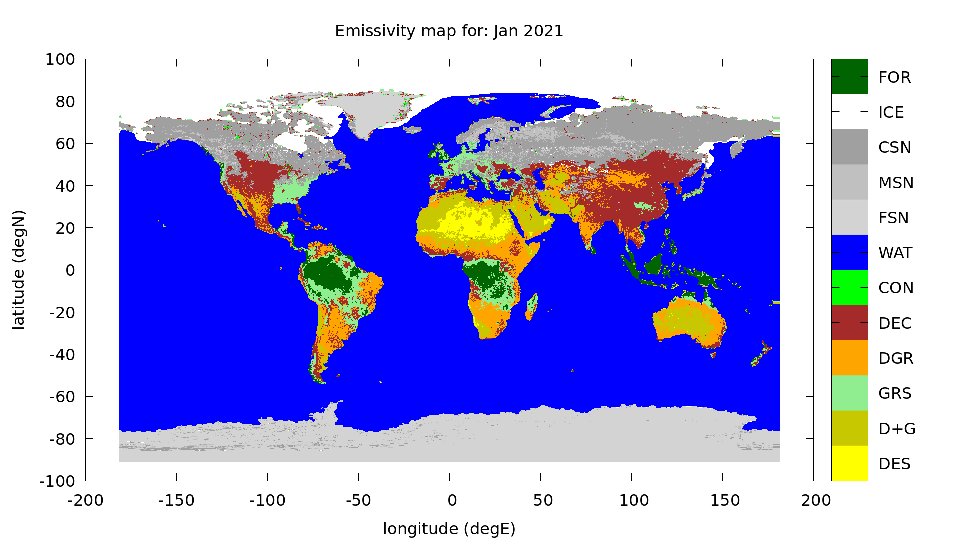}
\caption{Map of selected Huang profile for January 2021}\label{fig:method1_jan}
\end{figure}

\begin{figure}[ht!]
\centering
\includegraphics[width=1.
\textwidth]{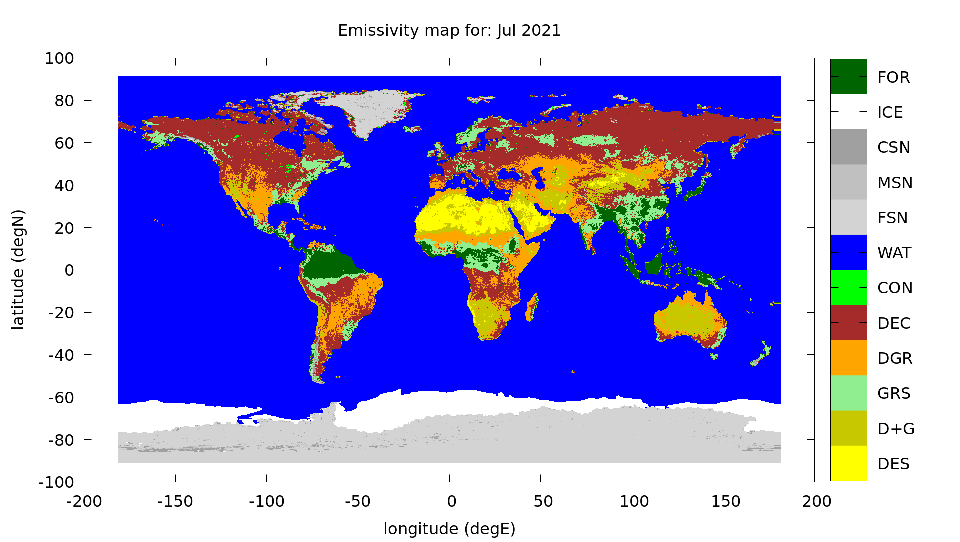}
\caption{Map of selected Huang profile for July 2021}\label{fig:method1_jul}
\end{figure}

The ancillary data that will be used as constrains are:
\begin{enumerate}
\item The snow cover. If the fraction is larger than a threshold, then snow/ice profiles instead of land profiles will be used. We set the threshold to $50$\%. The snow cover is only available over land.
\item The surface temperature. This will be used to impose some restrictions on the profiles, but mainly to distinguish pure water from ice on geolocations that are located over the oceans. While the freezing point of ocean water starts at $-1.8^\circ$~\si{C}, the fraction and depth of ice depends on how long and how lower than the freezing point the temperature remains \cite{seaice}. Since on the sea we have no emissivity data, we set a threshold value of $-6^\circ$~\si{C} to distinguish water from ice. Also, some mild constrains are used for some particular type of soil.
\item The soil humidity. Here again some empirical considerations were used to exclude arid soil types if the humidity is high. Also, the forest profile is expected to have a high humidity. 
\end{enumerate}
The full list of constrains is given is Table~\ref{ta:emissivity}. Again, we stress that this choice only determines the a-priori for the emissivity profile, the retrieved profile is determined by inverting the radiative transfer equation.
\begin{table}
\begin{center}
\begin{tabular}{ |l|c|c|c|c| } 
  \hline
  \textbf{ACR} & \textbf{SOIL TYPE} & \textbf{SNOW} &\textbf{TSKIN} & \textbf{HUMID}\\ 
  \hline
  DES & Desert & N & $\geq 20^\circ$\unit{C}& $\leq 20$\%\\
  \hline
  D\&G & Desert 45\%  and grass 55\% & N & $\geq 0^\circ$\unit{C} & $\leq 25$\%\\
  \hline
  GRS & Grass & N & --- & $10$\%-$45$\%\\
  \hline
  DGR & Dry grass & N & --- & $\leq 35$\%\\
  \hline
  DEC &  Deciduous & N & --- & $10$\%-$35$\%\\
  \hline
  CON & Conifer & N & --- & $20$\%-$45$\%\\
  \hline
  WAT & Water & N & $\geq -6^\circ$\unit{C} & --- \\
  \hline
  FSN & Fine snow & Y & --- & --- \\
  \hline
  MSN & Medium snow & Y & --- & ---\\
  \hline
  CSN & Coarse snow & Y & --- & ---\\
  \hline
  ICE & Ice & Y & $\leq -6^\circ$\unit{C} & --- \\
  \hline
  FOR & Forest & N & $\geq 4^\circ$\unit{C} & $\geq 40$\% \\
  \hline
\end{tabular}
\caption{List of acronyms for Huang emissivity profiles, and ancillary data constraints used. Snow cover larger than $50$\% flag, surface temperature and humidity admissible ranges.}\label{ta:emissivity}
\end{center}
\end{table}
For each ERA5 grid point, we select the soil types that do not violate the constraints reported in Table~\ref{ta:emissivity}, and then choose within this set the one with the minimal Root Mean Square (RMS) error in emissivity with respect to the MODIS database. On water, when there is no MODIS data, the profile is decided by the surface temperature. The results are shown in Figure~\ref{fig:method1_jan} for January 2021, and in Figure~\ref{fig:method1_jul} for July 2021.

Both maps are defined in the $[-85^\circ,85^\circ]$ latitude range, so the Summer polar North ice cap does not appear. Note that the snow covering is the only parameter linked to the soil type that is used in this identification.

\begin{figure}[ht!]
\centering
\includegraphics[width=1.
\textwidth]{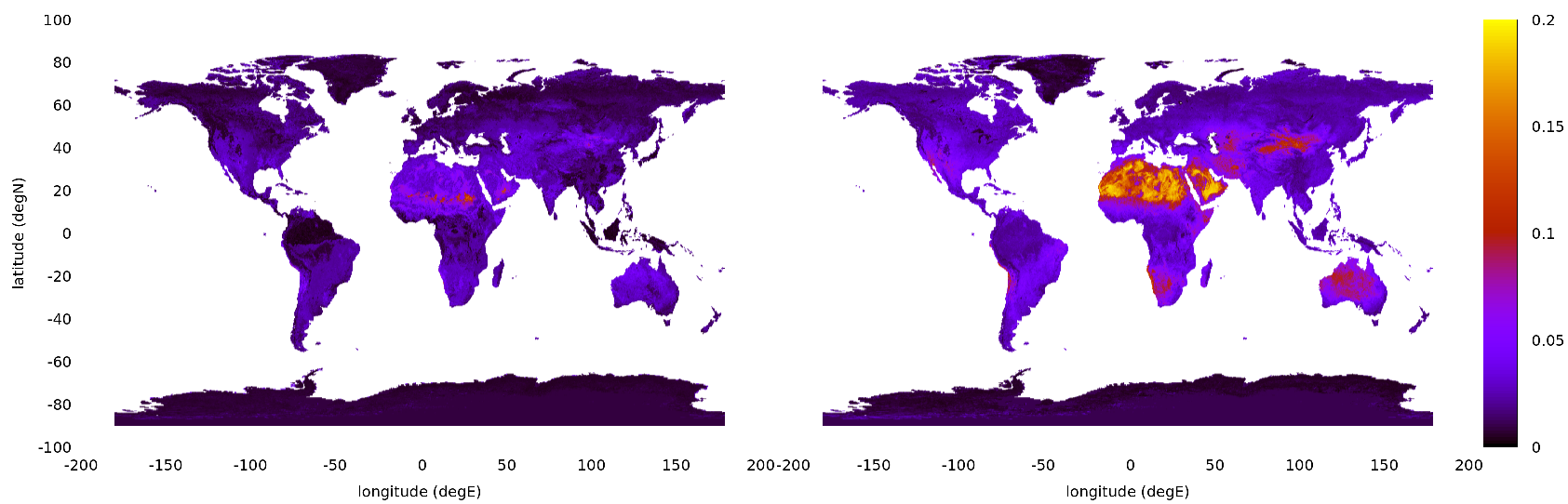}
\caption{RMS error with respect to MODIS data for July 2021. Left panel (selected Huang profiles), right panel (constant emissivity $0.99$)}\label{fig:error_jul}
\end{figure}

In Figure~\ref{fig:error_jul} we report the average RMS error with respect to the MODIS data in July. A similar plot holds also for January. In the left panel we show the error in the case of the selected Huang profile, in the right panel the error in case a constant emissivity equal to $0.99$ is selected. From the figure we note that selecting an appropriate Huang profile reduces the initial error as expected. Even in the case of the Huang profiles, there is a residual noticeable error mostly concentrated in the arid zones. This is due to the fact that there is a large variability in the desert emissivity, due to the presence of the quartz Reststrahlen bands in the sandy areas \cite{salisbury1992,masiello2014} but not in the rocky areas. We also repeated the method using the CAMEL data. The results are very similar, so we skip presenting them.

\subsection{Selection of profiles via Bayesian approach}\label{subse:bayesia}
The Huang profiles are given on the $\nu_k=50+5(k-1)$ set of wavenumbers, $k=1,...,321$. This set covers the $[50,1650]$~\unit{cm^{-1}} band at a resolution of $5$~\unit{cm^{-1}}, an interval sufficient to calculate the simulated FORUM spectra. In this approach we search an optimal estimation profile $e(\nu)$ as a convex combination of the Huang profiles $H=\{H_i(\nu)$, $i=1,...,12\}$. 
\begin{equation}\label{eq:optimal_def}
e(\nu)=\sum_{i=1}^{12}p_iH_i(\nu),\quad p_i\geq 0,\quad \sum_{i=1}^{12}p_i=1
\end{equation}
We define an a-priori for the emissivity making use of the MODIS/Terra+Aqua Yearly Land Cover Type database in \cite{Friedl2015}, that contains a static estimate of the type of soil with a $0.05^\circ$ grid in latitude and longitude. The 17 soil types considered are different from those considered by the Huang profiles, they are listed in Table~\ref{tab:landcover}.
\begin{table}
\begin{center}
\addtolength{\tabcolsep}{-0.35em}
\begin{tabular}{|r|l|r|l|}
\hline
1&barren or sparsely vegetated&
2&snow and ice\\
\hline
3&cropland/natural vegetation mosaic&
4&urban and built-up\\
\hline
5&croplands&
6&permanent wetlands\\
\hline
7&grasslands&
8&savannas\\
\hline
9&woody savannas&
10&open shrublands\\
\hline
11&closed shrubland&
12&mixed forests\\
\hline
13&deciduous broadleaf forest&
14&deciduous needleleaf forest\\
\hline
15&evergreen broadleaf forest&
16&evergreen needleleaf forest\\
\hline
17&water&&\\
\hline
\end{tabular}
\end{center}
\caption{Definition of the 17 land cover types.}\label{tab:landcover}
\end{table}

Thus, we established a correspondence $\bfM$ between the soil types of the land cover map and the Huang profiles. Let $T=\{T_l,\ l=1,...,17\}$ the soil types of the database in \cite{Friedl2015}, we define an equivalence mapping from the $T$ set of soils to the $H$ set of emissivity profiles. Let $\bfM=(m_{ij})$ be the $17\times 12$ matrix such that for each MODIS soil type $i$ we assign a probability $m_{ij}$ for each Huang profile $j$. Then $m_{ij}\geq 0$, and $\sum_j m_{ij}=1$ for every $i$. The actual matrix $\bfM$ is shown in Table~\ref{tab:correspondence}. Then, for each ERA5 grid point we calculated the probability of each of the $17$ soil types by averaging the points that would be enclosed in the FORUM field of view, if the center of the FORUM pixel were the ERA5 grid point, i.e. points with a distance less of equal $7.5$~\unit{km} from the center. In this way we obtain a sample of the emissivities in the whole globe with the same grid of the ERA5 data.

\begin{table}
\begin{center}
\addtolength{\tabcolsep}{-0.35em}
\begin{tabular}{|r|r|r|r|r|r|r|r|r|r|r|r|r|}
\hline
\textbf{LT}&DES&D+G&GRS&DGR&DEC&CON&WAT&FSN&MSN&CSN&ICE&FOR\\
\hline
\hline
1&0.5&0.3&0&0.2&0&0&0&0&0&0&0&0\\
\hline
2&0&0&0&0&0&0&0&0.25&0.25&0.25&0.25&0\\
\hline
3&0&0&0.5&0&0.5&0&0&0&0&0&0&0\\
\hline
4&0&0&0.1&0.1&0.8&0&0&0&0&0&0&0\\
\hline
5&0&0&0.1&0&0.9&0&0&0&0&0&0&0\\
\hline
6&0&0&0.2&0&0&0&0.8&0&0&0&0&0\\
\hline
7&0&0&0.8&0&0.1&0&0&0&0&0&0&0.1\\
\hline
8&0&0.2&0.4&0.2&0.2&0&0&0&0&0&0&0\\
\hline
9&0&0&0.3&0&0.3&0&0&0&0&0&0&0.4\\
\hline
10&0&0&0&0&0.6&0.2&0&0&0&0&0&0.2\\
\hline
11&0&0&0&0&0.4&0.3&0&0&0&0&0&0.3\\
\hline
12&0&0&0&0&0.2&0.4&0&0&0&0&0&0.4\\
\hline
13&0&0&0&0&0.5&0&0&0&0&0&0&0.5\\
\hline
14&0&0&0&0&0.5&0.5&0&0&0&0&0&0\\
\hline
15&0&0&0&0&0&0&0&0&0&0&0&1\\
\hline
16&0&0&0&0&0&1&0&0&0&0&0&0\\
\hline
17&0&0&0&0&0&0&1&0&0&0&0&0\\
\hline
\end{tabular}
\end{center}
\caption{Correspondence between Land Cover Map and Huang profiles.}\label{tab:correspondence}
\end{table}

Then, if $t_l$ are the fractions of the $T_l$ soils in the FORUM pixel, we can find an a-priori emissivity profiles by
defining:

\begin{equation}\label{eq:huang_ap}
H_a(\nu)=\sum_{l=1}^{17}t_l \sum_{j=1}^{12}m_{lj}H_{j}(\nu).
\end{equation}

We can group the coefficients of the Huang profiles,  defining $a_i=\sum_{l=1}^{17} t_lm_{li}$. Then, (\ref{eq:huang_ap}) can be recast as:

\begin{equation}\label{eq:huang_ap2}
H_a(\nu)=\sum_{i=1}^{12}a_i H_{i}(\nu),
\end{equation}
where $a_i\geq 0$, $\sum_i a_i=1$.

The optimal emissivity profile should not be far from the a-priori. However, the optimal emissivity profile should also be compatible with the chosen emissivity database. For this experiment we selected the CAMEL database. Let $n_j$ and $e_j^{(k)}$ $j=1,...,9$ the CAMEL wavenumbers and monthly emissivity values for an ERA5 grid point $k$ that are inside the FORUM spectrum. Let $\bar{e}_j$ the average value of the $e_j^{(k)}$, then:
\begin{equation}\label{eq:Sm}
{\bfS_C}_{ij}=\frac{1}{N}\sum_{k\in {\rm ERA5\ grid}} (e_i^{(k)}-\bar{e}_i)(e_j^{(k)}-\bar{e}_j)
\end{equation}
is the experimental VCM of the CAMEL emissivities.

Let $\bar{h}_j$ be the average of the $12$ Huang profiles for the wavenumber $\nu_j$, $j=1,...,321$. We can calculate the Huang VCM as:
\begin{equation}\label{eq:Sm1}
{\bfS_H}_{ij}=\frac{1}{12}\sum_{k=1}^{12}(H_k{(\nu_i})-\bar{h}_i)(H_k{(\nu_j})-\bar{h}_j).
\end{equation}
However, this $321\times 321$ matrix does not really represent the real variability of the Huang profiles, because the correlation between adjacent points is very large. Thus, we use the technique of McMillin and Goldberg \cite{mcmillin1997}, also used by Matricardi \cite{matricardi2010} for the IASI (Infrared Atmospheric Sounding Interferometer) sounder, to reduce the cardinality of the VCM. We first select the wavenumber $\bar{\nu}_1=\nu_{j_1}$ as the one with the largest standard deviation, i.e. ${\bfS_H}_{j_1j_1}\geq{\bfS_H}_{ii}$, $\forall i$. Then we discard all wavenumbers $\nu_l$ such that:
\begin{equation}\label{eq:correlation_remove}
|{\bfS_H}_{j_1l}|\geq c\ \sqrt{{\bfS_H}_{j_1j_1}{\bfS_H}_{ll}}
\end{equation}
for some constant $0<c<1$ and repeat the procedure until all additional wavenumbers are removed. In this way we select a subset of wavenumbers $\nu_{j_l}$, $l=1,...,L$ (the \emph{super channels} in the original terminology) that is representative of the variability of the Huang profiles. 

\begin{figure}[ht!]
\centering
\includegraphics[width=1.
\textwidth]{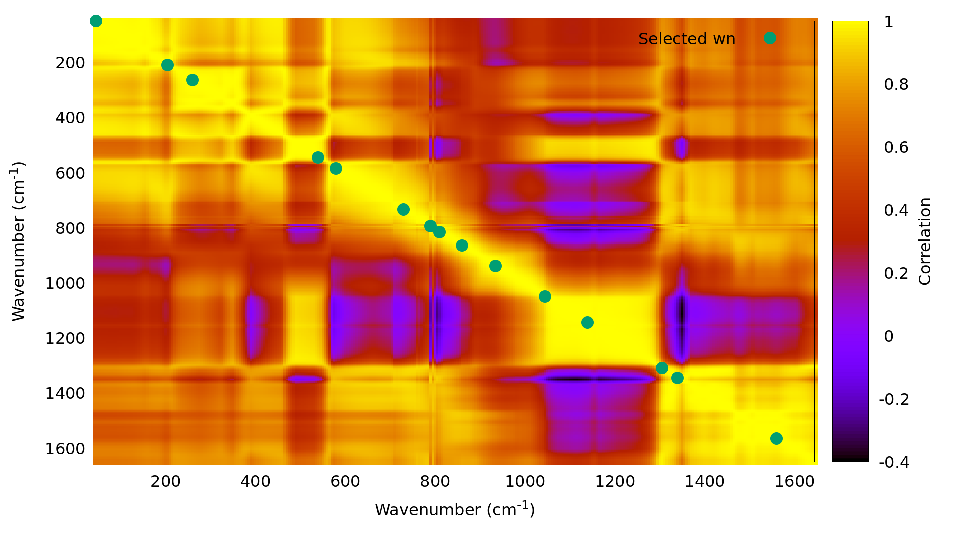}
\caption{Correlation matrix for the set of Huang profiles and selected wavenumbers}\label{fig:vcm_sel}
\end{figure}

The result of the procedure with $c=0.9$ is reported in Figure~\ref{fig:vcm_sel}, where we obtain $L=15$.
Let us define ${\bfS_R}$ the $L\times L$ VCM obtained taking the set of $j_l$ rows and columns from ${\bfS_H}$. Thus, if we suppose that both the difference from the CAMEL data and from the a-priori are random Gaussian variables, the combined probability of finding an optimal profile $e^{(k)}(\nu)\equiv\sum_i p^{(k)}_i H_i(\nu)$ for the ERA5 grid point $k$ is:
\begin{equation}\label{eq:correlation}
\begin{split}
p(e^{(k)}(\nu))&=\frac{\det(\bfS_C)^{-1/2}}{(2\pi)^{-9/2}}\exp\left(-\frac{1}{2}(\bfe^{(k)}(\bfn_C)-\bfe_C^{(k)})^t{\bfS_C^{-1}}(\bfe^{(k)}(\bfn_C)-\bfe_C^{(k)})\right)\\
&\cdot \frac{\det(\bfS_R)^{-1/2}}{(2\pi)^{-L/2}}\exp\left(-\frac{1}{2}(\bfe^{(k)}(\bfn_R)-\bfe_R^{(k)})^t{\bfS_R^{-1}}(\bfe^{(k)}(\bfn_R)-\bfe_R^{(k)})\right),
\end{split}
\end{equation}
where $\bfn_C$ is the vector of the $9$ CAMEL wavenumbers $n_j$ included in the FORUM spectral range, $\bfn_R$ is the vector of the $L$ reduced set of Huang wavenumbers $\nu_{j_l}$, $\bfe_C^{(k)}$ is the vector of CAMEL emissivities for the $k$ ERA5 grid point, $\bfe_R^{(k)}$ is the vector of the a-priori Huang emissivites $H_a$ on the $\bfn_R$ vector, and finally $\bfe^{(k)}(\bfn_C)$ and $\bfe^{(k)}(\bfn_R)$ are the values of the $e^{(k)}(\nu)$ function defined in (\ref{eq:optimal_def}) on the vectors $\bfn_C$ and $\bfn_R,$ respectively. The probability of (\ref{eq:correlation}) can be maximized by minimizing the sum of the exponents:
\begin{equation}\label{eq:bayes}
\begin{split}J(p_1,...,p_{12})&=(\bfe(\bfn_C)-\bfe_C)^t{\bfS_C^{-1}}(\bfe(\bfn_C)-\bfe_C)\\
&+(\bfe(\bfn_R)-\bfe_R)^t{\bfS_R^{-1}}(\bfe(\bfn_R)-\bfe_R),
\end{split}
\end{equation}
where for simplicity we dropped the dependence from $k$ in the notation. The functional (\ref{eq:bayes}) has the obvious constrains $p_i\ge 0$, $\sum_ip_i=1$, see (\ref{eq:optimal_def}). If the minimization were carried out on all $p_i$, the minimization would insert profiles that are not compatible with the geolocation in order to minimize the distance from the CAMEL data, for instance snow profiles in the tropics. To avoid this problem, the minimization is carried out only on the Huang profiles contained in the a-priori, that is profiles $i$ such that $a_i>0$ in (\ref{eq:huang_ap2}). The functional $J$ is a quadratic function of the $p_i$, so (\ref{eq:bayes}) can be solved explicitly with the constrained linear least-square method.

\section{Test of the method on experimental IASI data}
The IASI (Infrared Atmospheric Sounding Interferometer) sounder \cite{clerbaux2008} is a Fourier transform interferometer that measured atmospheric spectra since 2006. The IASI interferometer swath includes $14$ Fields of Regard (FoR). Each FoR consists of $4$ Fields of View (FoV) that are acquired simultaneously. Each FoV corresponds to a spectrum. The different FoRs have different nadir angles. The land covered by each swath is a strip about $2000$~\unit{km} wide perpendicular to the IASI orbit. To avoid the influence of the observation angle that distorts the shape of the FoV, we only selected the two central FoRs, which have an inclination not larger than $8$ degrees with respect to the nadir. The shape of these FoVs is roughly circular with a $12$~\unit{km} diameter. Moreover, we discarded FoVs over water and FoVs that are considered cloudy. The IASI FoVs considered have their own geolocation, which can be read from the IASI Level 2 files.

For the comparison tests we used the emissivity data available in the IASI Level 2 products provided by EUMETSAT Data Services \cite{iasilevel22020}, which are available online from the site: \url{https://pn-ui.prod.wekeo2.eu/pn-ui/start}. 
The emissivity values in the IASI products are given for the following channels: $765$, $900$, $991$, $1071$, $1160$, $1228$, $1980$, $2111$, $2170$, $2510$, $2634$, and $2760$~\unit{cm^{-1}}. We only consider the first $6$ wavenumbers, as they fall into the FORUM spectral range.

To conduct the tests consistently with the datasets used in the generation of the ERA5 grid, we selected orbits from January and July 2021; specifically, we considered the orbits from July 15th and 16th, and some orbits from January 16th, 19th, and 20th. To determine coincidences between the ERA5 grid, on which both the CAMEL and the Bayesian profiles are based, and the IASI FoVs, we set a tolerance of $\pm0.05^\circ$ for longitude and latitude between the center of the IASI FoV and the ERA5 grid point. Using these criteria, we obtained a sample of $4530$ coincidences for January and $4480$ coincidences for July.

To assess the Bayesian test, we compared the Root Mean Square Error (RMSE) of the difference between the Bayesian test (labelled BAYES) and the IASI data, and the RMSE of the difference between the CAMEL profile, obtained by interpolating the CAMEL values, and the IASI data.

\begin{figure}[ht!]
\centering
\includegraphics[clip,trim={0 105 0 130}, width=1.
\textwidth]{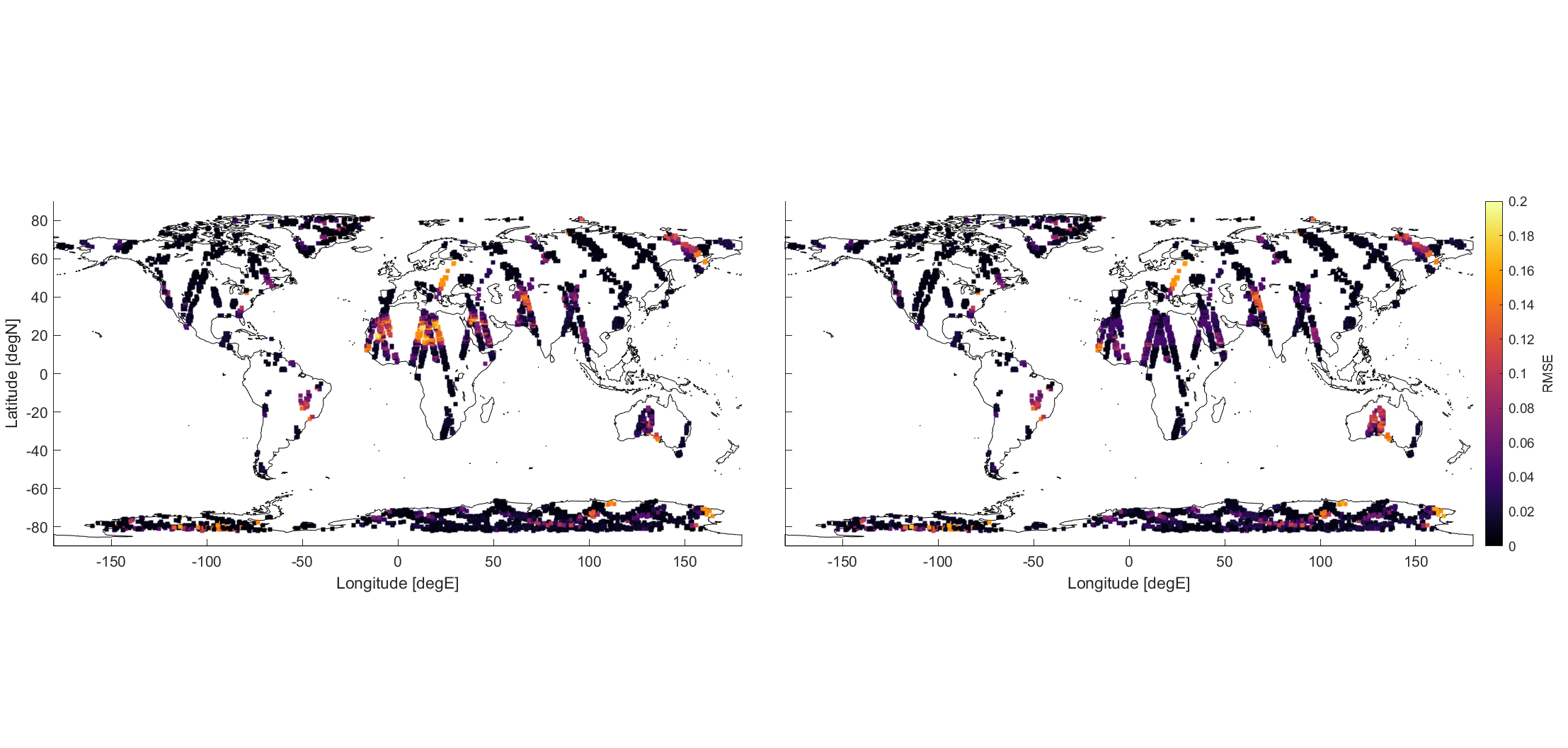}
\caption{RMSE of the emissivity for the January month. Left panel CAMEL-IASI, right panel Bayesian-IASI}\label{fig:rms_january}
\end{figure}

\begin{figure}[ht!]
\centering
\includegraphics[clip,trim={0 105 0 130}, width=1.
\textwidth]{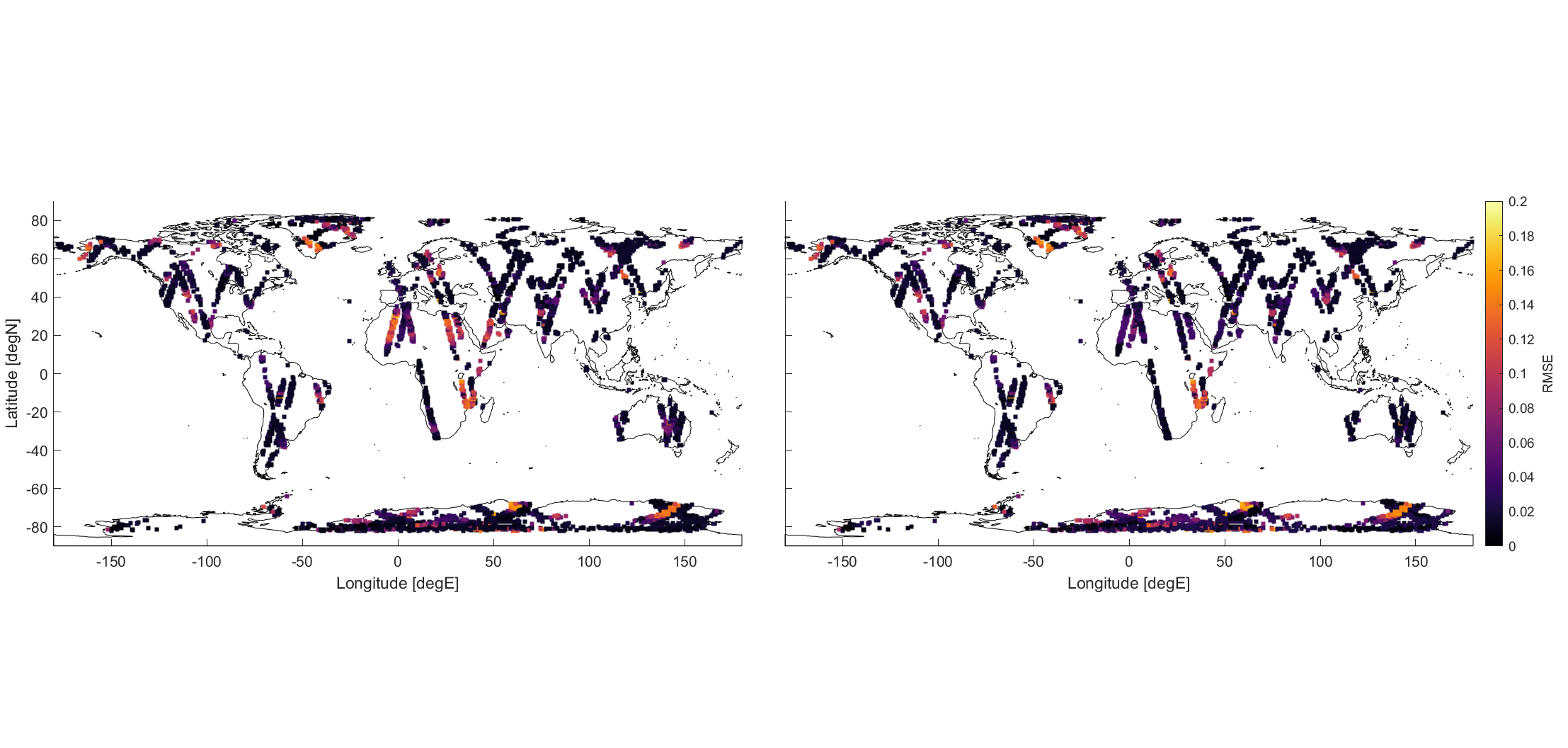}
\caption{RMSE of the emissivity for the July month. Left panel CAMEL-IASI, right panel Bayesian-IASI}\label{fig:rms_july}
\end{figure}

In Figures~\ref{fig:rms_january} and \ref{fig:rms_july} we show the results of the test, respectively for the January and July month. We note that the Bayesian test agrees better with the IASI data over Africa for both tests, while there is a worsening over Australia in the January test.

Table~\ref{ta:results} synthetizes the results, reporting the average RMSE between the tests, calculated on the samples. From the second row we note that the IASI and CAMEL data are not fully compatible. This has to be expected, since the CAMEL data are monthly mean on a grid with $0.25^\circ$ step, while the IASI data refers to a FoV that is $12$~\unit{km} in diameter. We note that our Bayesian method finds a solution which is not far from the CAMEL data, but agrees better with the experimental IASI data. We attribute this result to the fact that, while the CAMEL solution does not include small-scale features that could not be deduced from the CAMEL sampling, the Bayesian solution is determined as a convex combination of high-resolution profiles that contains small-scale features of soil types that are validated by measurements.

\begin{table}
\begin{center}
\begin{tabular}{ |l|c|c| } 
  \hline
 & \textbf{January} & \textbf{July} \\ 
  \hline
  \text{BAYES-IASI} & 0.0264 & 0.0267 \\
  \hline
  \text{CAMEL-IASI} & 0.0297 & 0.0283 \\
  \hline
  \text{CAMEL-BAYES} & 0.0191 & 0.0150 \\
  \hline
\end{tabular}
\caption{Average RMSE of the differences between IASI data, the CAMEL profiles and the BAYES solution.}\label{ta:results}
\end{center}
\end{table}

Finally, we calculated the significance of the deviation from the CAMEL data of the BAYES solution, using a two-sample t-test. The decrease in the RMSE is significant for both the January and the July test. For the January test the probability that the CAMEL sample is drawn from a distribution with an average RMSE lower or equal that of the BAYES sample is $0.000006$. For the July test the value is $0.026926$.  

\section{Conclusions}
In this paper we studied the problem of selecting an emissivity profile that is as close as possible to the experimental data for that geolocation and time. This is important in the retrieval step, because in the radiative transfer equation the dependence of emissivity and surface temperature is very similar, so that assessing both quantities may lead to biases.

We tested two methods for determining an emissivity profile for each geolocation. The first one only uses MODIS and ancillary data that do not include any land cover information, but is still able to obtain a plausible soil type map. The second method is a Bayesian approach that uses the CAMEL database and a detailed land cover map to obtain an emissivity profile in the form of a convex combination of the Huang profiles.

The main result of this paper is that, in the test with IASI experimental data, the Bayesian method is better than the linear spline obtained by interpolating the CAMEL measurements. This is because the Huang profiles are able to recover the small-scale features of the spectral emissivity that the linear spline, which nevertheless reconstructs the CAMEL values exactly, is not able to capture.

From the point of view of applications, the Bayesian method for determining the emissivity profile is not computationally time-demanding; the generation of a single case profile takes milliseconds on a standard PC. In the future, using this method to determine the initial guess and/or the a-priori for the emissivity should reduce the bias in the retrieval of surface parameters.

\section*{CRediT authorship contribution statement}
{\bf Luca Sgheri:} Methodology, Formal Analysis, Software, Writing – original draft, Writing – review \& editing, Funding Acquisition; {\bf Cristina Sgattoni:} Formal Analysis, Software, Writing – original draft, Writing – review \& editing; {\bf Chiara Zugarini:} Formal Analysis, Software, Writing – original draft, Writing – review \& editing; 

\section*{Acknowledgements}
The paper was supported by the CNR project DIT.AD012.168 Support to the FORUM mission. The authors wish to thank INdAM mathematical research groups GNCS and GNAMPA for support.
\bibliographystyle{model1b-num-names}
\bibliography{bibliography}

\end{document}